# Skull Flexure from Blast Waves: A Mechanism for Brain Injury with Implications for Helmet Design[*]

William C. Moss[1], Michael J. King[1], and Eric G. Blackman[2]

[1]*Lawrence Livermore National Laboratory, Livermore, CA 94551*

[2]*Department of Physics and Astronomy, University of Rochester, Rochester NY 14627*

## Abstract

Traumatic brain injury [TBI] has become a signature injury of current military conflicts, with debilitating, costly, and long-lasting effects. Although mechanisms by which head impacts cause TBI have been well-researched, the mechanisms by which blasts cause TBI are not understood. From numerical hydrodynamic simulations, we have discovered that non-lethal blasts can induce sufficient skull flexure to generate potentially damaging loads in the brain, even without a head impact. The possibility that this mechanism may contribute to TBI has implications for injury diagnosis and armor design.

[*] *NOTICE: This article has been authored by Lawrence Livermore National Security, LLC ("LLNS") under Contract No. DE-AC52-07NA27344 with the U.S. Department of Energy for the operation of Lawrence Livermore National Laboratory ("LLNL"). Accordingly, the United States Government retains and the publisher, by accepting the article for publication, acknowledges that the United States Government retains a nonexclusive, paid-up, irrevocable, world-wide license to publish or reproduce the published form of this article, or allow others to do so for United States Government purposes.*

Traumatic brain injury [TBI] results from mechanical loads in the brain, often without skull fracture, and causes complex, long lasting symptoms (*1,2*). TBI in civilians is usually caused by head impacts resulting from motor vehicle (*3,4*) and sports accidents (*5,6*). TBI has also emerged to be endemic among military combat personnel exposed to blasts. As modern body armor has substantially reduced soldier fatalities from explosive attacks, the lower mortality rates have revealed the high prevalence of TBI (*1,7,8*). There is an urgent need to understand the mechanisms by which blasts cause TBI, to better diagnose injury and design protective equipment, such as helmets.

Impact-induced TBI [ITBI] has been extensively studied, primarily through animal testing and analyses of human trauma data (*9*), and has been linked to accelerations of the head. By contrast, the damage producing mechanisms for blast-induced TBI [BTBI] are not well understood (*10,11*). Mechanical loads from the blast pressure, accelerations, or impacts, as well as electromagnetic or thermal exposure have all been proposed (*12*). Because blasts can cause head impacts by propelling a soldier into another object (or vice versa), protection research has traditionally focused on reducing the acceleration of the head during an impact. However, shock tube experiments in which restrained animals were subjected to blast-like conditions confirmed that blast pressures, without subsequent impacts, can cause TBI (*13*). Several mechanisms by which the blast alone can damage the brain have been proposed, including bulk acceleration of the head (*12*), transmission of loads through orifices in the skull, and compression of the thorax, which generates a vascular surge to the brain (*13*). Surprisingly, blast-induced deformation of the skull has been neglected, perhaps due to the perception that the hard skull protects the brain from non-lethal blast waves (*14*). Here we show via three-dimensional hydrodynamical simulations that direct action of the blast wave on the head causes skull flexure, producing mechanical loads in brain tissue comparable to those in an injury-inducing impact, even at non-lethal blast pressures as low as 1 bar above ambient.

We studied head impacts and blast waves on the head using ALE3D (*15*), an arbitrary Lagrangian-Eulerian [ALE] finite element hydrocode. Figure 1 shows our blast simulation geometry. The charge size and standoff distance from the simulated head were chosen to generate a non-lethal blast wave (*16*). The skull is modeled as a hollow elastic ellipsoid that contains a viscoelastic brain surrounded by a layer of cerebrospinal fluid [CSF]. The tensile stress that the CSF layer can carry is capped at one bar below atmospheric pressure to capture cavitation-like effects (*17,18*), although it is not clear if the CSF itself cavitates due to the presence of impurities and dissolved gas (*19*), or if the interfaces between the CSF and the subarachnoid walls cannot support tensile stresses. Because the CSF layer is thin, capping its tensile strength models either scenario. A simplified face (with no lower jaw), neck, and body are included to capture blast-induced accelerations accurately, and to appropriately shield the bottom of the braincase from the blast wave. Anatomical details such as skull thickness variations, grey/white matter, ventricles, etc. are not included. Although these features are needed to predict specific medical traumas, our simplified model quantitatively distinguishes the different mechanisms by which impacts versus blasts load the brain. It also provides a means of exploring protective strategies: a helmet that reduces the magnitude of these loads would necessarily reduce TBI.

For our impact simulations we encased the head model described above in a steel-shelled helmet containing an inner layer of crushable foam, as shown in Figure 2a. The head and helmet were impacted against a rigid wall. We chose impact velocity and foam parameters to produce an acceleration load consistent with typical ITBI, according to the commonly used Head Injury Criterion [HIC] measure (*20*), which derives from empirical data of automotive crash tests (*9*). For our choice of foam and an impact velocity of 5 m/s, the average acceleration was 194 G's for 2.1 ms. This corresponds to an HIC = 1090, comparable to the motor vehicle injury standard of 1000.

Our impact simulations revealed known mechanisms of ITBI (*6*). Figure 2a shows the brain pressures at the moment of maximal deceleration. The brain collides with the decelerating skull and develops large positive pressure at the "coup" and negative pressure at the "contrecoup." The rebound of the brain then creates pressure spikes, pressure gradients, and shear strains at the contrecoup. The brain oscillates until the impact energy is dissipated. Because the head impacts the wall obliquely, it rotates and causes potentially damaging shear strains.

Blast simulation results for an unprotected head are shown in Figure 2b and Figure 2c, and indicate dramatically different loading modes acting on the brain than those resulting from impact. Figure 2b shows the pressure as the blast wave reaches the skull. It transits the body in ~0.7 ms at a speed of 450 m/s and an overpressure of 1 bar above ambient, inducing ~80 G's of bulk acceleration. Figure 2c shows an expanded view of the head with pressure contours in the air and brain, and velocity vectors in the skull. The moving pressure wave generates flexural ripples in the skull.

Skull flexure, not head acceleration, produces most of the mechanical load in the brain for the blast simulation. The skull is an elastic structure in contact with a deformable foundation (the CSF and brain). A concentrated load moving at high speeds (*i.e.* the blast wave front) over such a structure drives transverse bending displacements under and in front of the load (*21*). These displacements directly produce pressure extremes (0 to ~3 bar absolute pressure, neglecting high frequency transients) comparable to those in the ITBI simulations described above, and even larger pressure gradients (several bar/cm), because the extremes are closer together (compare Figure 2a and 2c). These loads occur despite a significantly smaller bulk acceleration and a shorter acceleration time compared to the impact simulations: the overpressure-induced acceleration only produces an HIC=18. The dominant role of skull flexure was confirmed by parametric studies where the skull stiffness was varied. For the same 1 bar blast, a skull 1000 times stiffer cut shear strains in half, peak pressure fivefold, and pressure gradients tenfold. Making the skull perfectly rigid and applying the same bulk accelerations as those generated by the blast resulted in even smaller loads.

We performed six additional simulations to confirm that our basic results were not sensitive to the geometry and symmetry of our skull model, or the mechanical properties of the brain, CSF, or skull. Using the simulation shown in Figures 1 and 2b-c as the base case, the following sensitivity studies were performed: (*i*) rotated the body and head 90°, to simulate a side-on blast; (*ii*) inserted holes into the skull to represent spinal column and optical nerve passages; (*iii*) increased the CSF layer tensile strength, to support arbitrarily large tensile loads; (*iv*) modified the material properties of the brain, reducing the bulk modulus and increasing the shear moduli and the viscoelastic decay rate (*22*); (*v*) increased just the shear moduli and the viscoelastic decay rate of the brain; (*vi*) replaced the elastic skull material with a viscoelastic material (*23*). Blast-induced skull flexure persists in all these variations. Cases (*i*) and (*ii*) produced no substantive differences from the base case (except for increased localized tissue shearing near the holes in case (*ii*)). Case (*vi*) produced no substantive difference during the first two milliseconds after the blast reaches the skull; at later times the skull's viscoelasticity damps the pressure oscillations. Shear strains in the brain, likely due primarily to head rotation, persist at late times regardless of the skull material.

Figure 3a compares pressure extrema in the brain as a function of time for the base case and cases (*iii*) and (*iv*). The transient pressure peak in the base case corresponds to the sudden recompression of the CSF layer near the front of the skull. Removing the tensile stress cap in case (*iii*) reduces the transient pressure peak by 25%; the elevated positive pressures due to localized skull flexure are otherwise identical. Additionally, hydrostatic tension greater than one bar below ambient develops in parts of the brain. These differences highlight the need to better characterize the effective *in vivo* tensile strengths of the CSF and its interfaces. However, the magnitude of the tensile strength has no effect on the occurrence of skull flexure.

The distinct features of case (*iv*) are due to the lower bulk modulus of the brain, because modifying only the shear properties (case (*v*)) produces nearly identical results to the base case. The peak skull displacements in case (*iv*) are the same as in the base case, resulting in generally lower peak pressures. The major difference between case (*iv*) and the base case is deeper penetration of pressure and pressure gradients into the brain, as shown in Figure 3b. This is likely due to the slow (~350 m/s) wave speed in the brain in case (*iv*), which does not allow the gradients to relax as quickly as in the base case, so the effects of localized flexure penetrate more deeply. There is significant variation in reported bulk moduli of brain tissue (*9*), especially when comparing *in vitro* and *in vivo* data. The sensitivity of the simulation results to the bulk modulus highlights a need for more accurate *in vivo* material characterization.

The specific paths by which mechanical loads in the brain lead to injury are still unknown (*7*), but we can speculate about how localized skull flexure might cause injury. Although we have modeled the brain as homogeneous, it is actually heterogeneous, with complex structures, interfaces and widely varying mechanical properties. When mechanical loads such as pressure waves or shear strains traverse material interfaces, amplified local shearing results, which is consistent with brain injuries such as diffuse axonal injury [DAI] being observed near material interfaces (*7*). In addition, pressure gradients across fluid-filled structures may mechanically damage these structures. Regardless of the specific mechanism, any TBI caused by external loads on the skull will be reduced if effective protective equipment reduces those loads.

We next studied how helmets and their suspension systems influence the blast-induced mechanical loads in the brain. We considered two common suspension systems that accommodate the ballistic standard of a 1.3 cm gap between helmet and head (*24*): a nylon web system, as formerly used in the Personnel Armor System Ground Troops [PASGT] infantry helmets, and viscoelastic foam pads like those in Advanced Combat Helmets [ACH]. The helmet was modeled as a hemi-ellipsoidal Kevlar shell in both cases.

Figure 4 is from a blast simulation of a helmet with a webbed suspension. The 1.3 cm gap allows the blast wave to wash under the helmet. When this "underwash" occurs, geometric focusing of the blast wave causes the pressures under the helmet to exceed those outside the helmet, so the helmet does not prevent the rippling deformation of the skull and the pressure gradients in the brain. For ACH-style foam-padded helmets, this underwash effect is mostly prevented, but motion of the helmet is more strongly coupled to the head. Helmet accelerations and bending deformations are transferred to the skull more effectively. The simulation results are very sensitive to the rate-dependent mechanical stiffness of the foam, which is not a well-measured quantity. Consequently, we varied the foam stiffness from values measured at low-rates to values three orders of magnitude larger. Foams that were stiffer at high loading rates transferred greater forces from the helmet to the skull and increased the mechanical loads in the brain relative to softer foams. But even soft foams only partially reduced the blast-induced pressures and pressure gradients in the brain, because the helmet does not cover enough of the head at the back and sides to prevent skull deformation.

In summary, we have provided evidence that the direct action of a non-lethal blast on the skull likely causes injury. Our simulations show that: (*i*) For a non-lethal blast with 1 bar of overpressure, accelerations imparted by the blast are likely too small to account for BTBI in the absence of other mechanisms; (*ii*) A blast wave causes the skull to dynamically deform, which creates localized regions of high and low pressure and large pressure gradients that sweep through the brain. Even modest skull flexure from a non-lethal blast wave produces loads at least as large as those from a typical injury-inducing impact; (*iii*) The localized skull flexure mechanism persists for different blast orientations, different effective tensile strengths of the CSF layer, different brain material properties, both elastic and viscoelastic skull properties, and in the presence of orifices in the skull. However, the pressure histories in the brain are sensitive to the brain bulk modulus and the effective tensile strength of the CSF layer; (*iv*) Helmets affect the interaction of the blast with the head. Without padding, the clearance gap between the helmet and the head allows underwash that amplifies pressures acting directly on the skull. Padding inhibits this underwash, but can more strongly couple helmet motion to the head, increasing the mechanical loads in the brain. If localized skull flexure proves to be a primary mechanism for BTBI, then an effective mitigation strategy would be to deny the blast wave access to the airspace under the helmet and prevent the motion and deformation of the helmet from transferring to the skull.

**Acknowledgments**: WCM thanks M. Moss for reading R. Glasser's article in the San Jose Mercury News [4-15-07] entitled "The Hidden Wounds of the Iraq War" and saying "You can simulate that, can't you?" We thank B. Watkins for support from the DoD/DOE Joint Munitions Program, M. Hale and S. Lisanby for discussions as part of the Defense Science Study Group of the Institute for Defense Analyses (EGB also a member), and T. Gay & T. Matula for reading the manuscript. This work performed under the auspices of the U.S. Department of Energy by Lawrence Livermore National Laboratory under Contract DE-AC52-07NA27344.

**Figures**

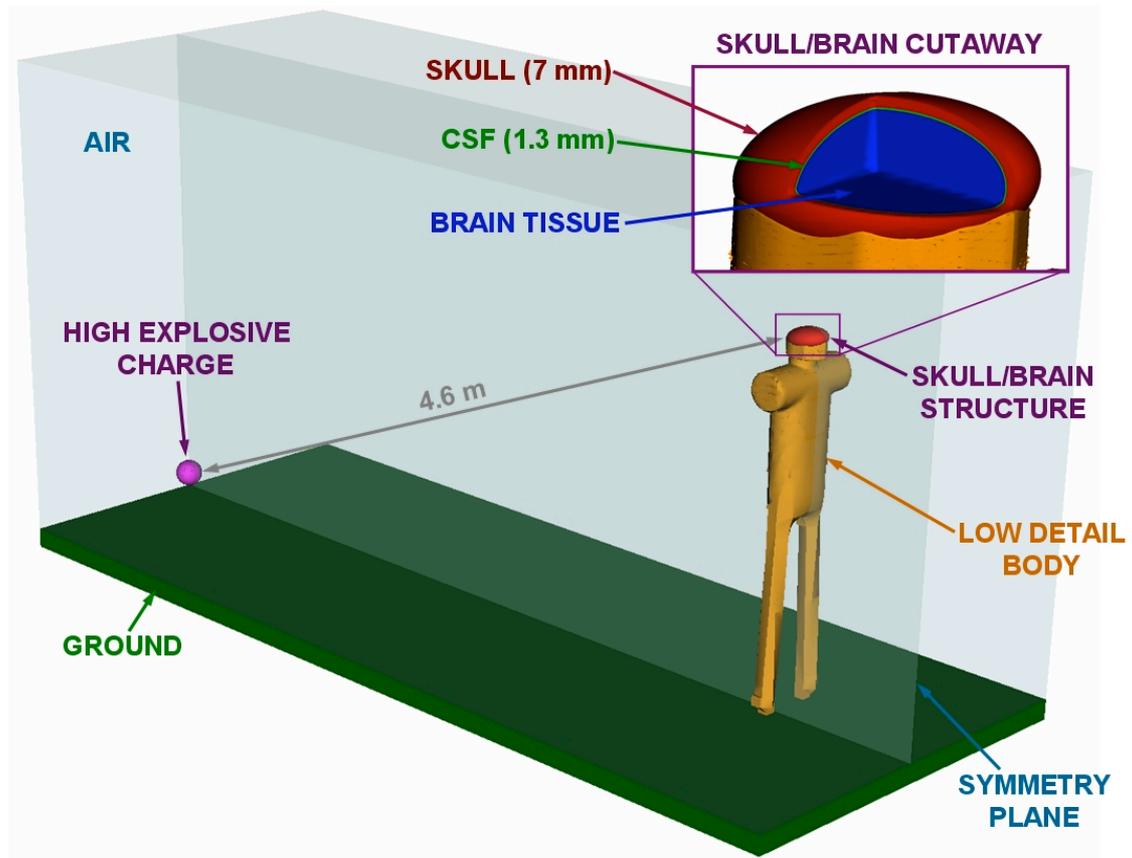

**Figure 1** – Simulation Geometry: A 2.3 kg spherical charge of C4 high explosive is located 4.6 m from a head consisting of three components—the skull, CSF layer, and brain tissue—that are supported by a low detail body structure.

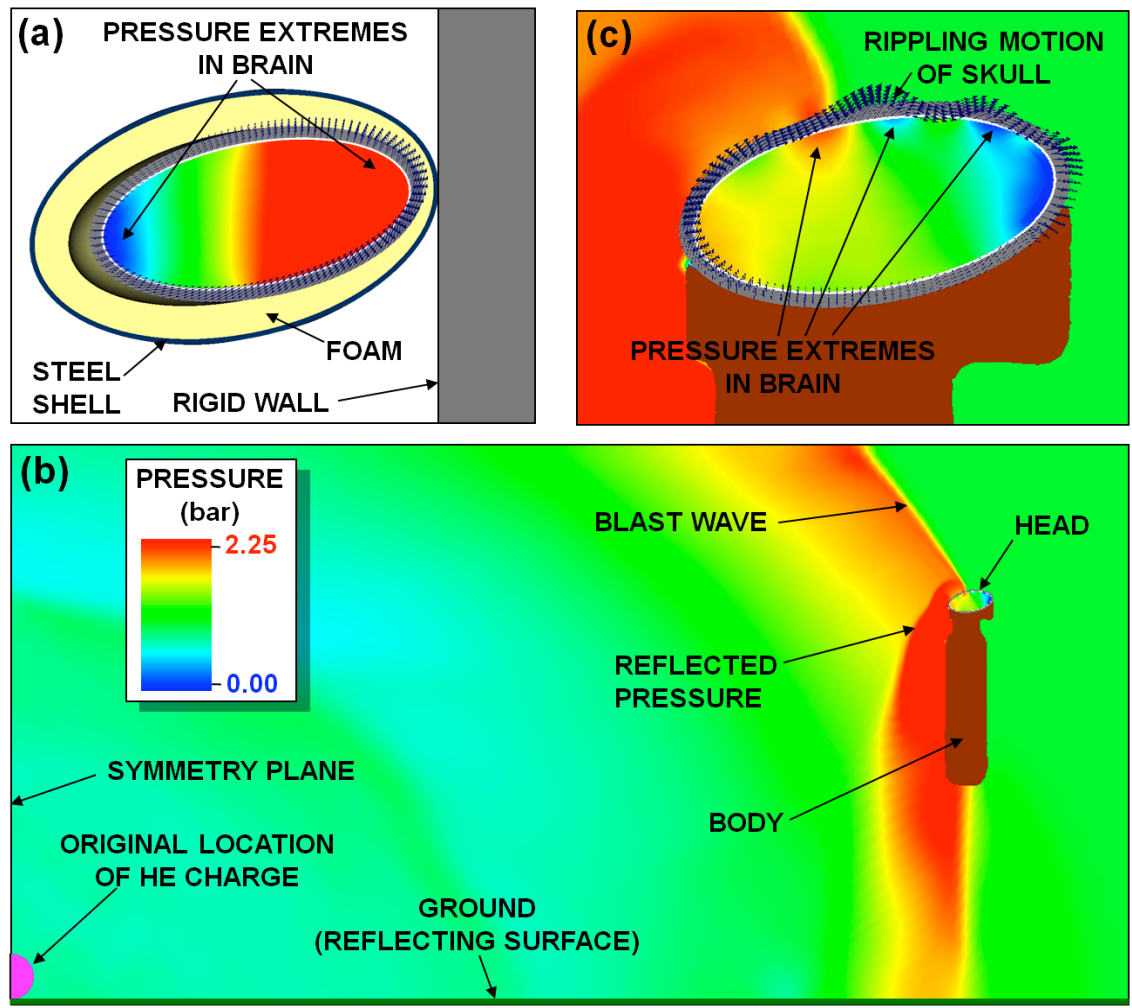

**Figure 2** – Pressure and skull motion for impact and blast simulations:

(a) Angled impact at maximum deceleration.

(b) Blast wave propagating past the simulated victim 5.6 ms after detonation.

(c) Expanded view of the head as the blast wave passes over it. Inward and outward rippling of the skull cause pressure extrema in the brain. The skull deflections are ~ 50um.

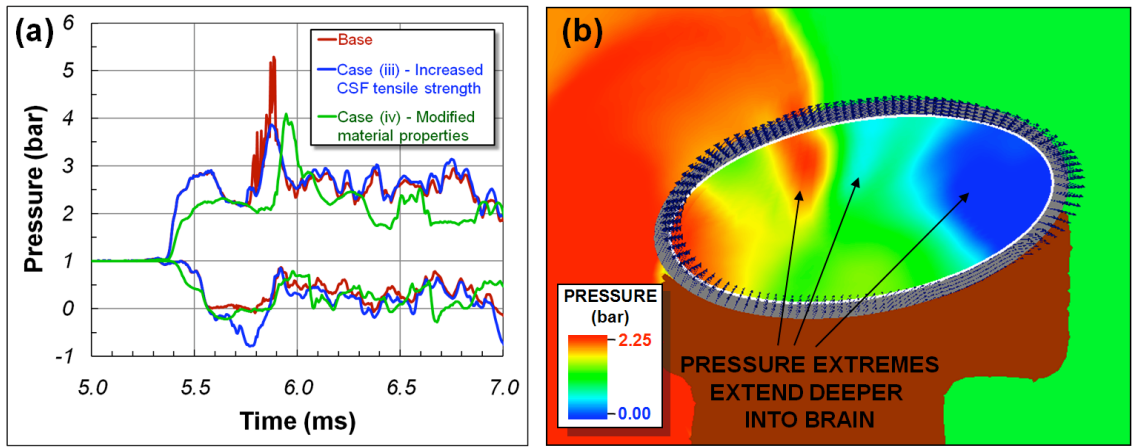

**Figure 3** – Selected results of sensitivity studies:

(a) Time history of maximum and minimum pressures occurring anywhere in the brain for the base case, case (*iii*), and case (*iv*).

(b) Pressure and skull motion for case (*iv*), 5.6 ms after detonation. Skull deflections are ~50 μm.

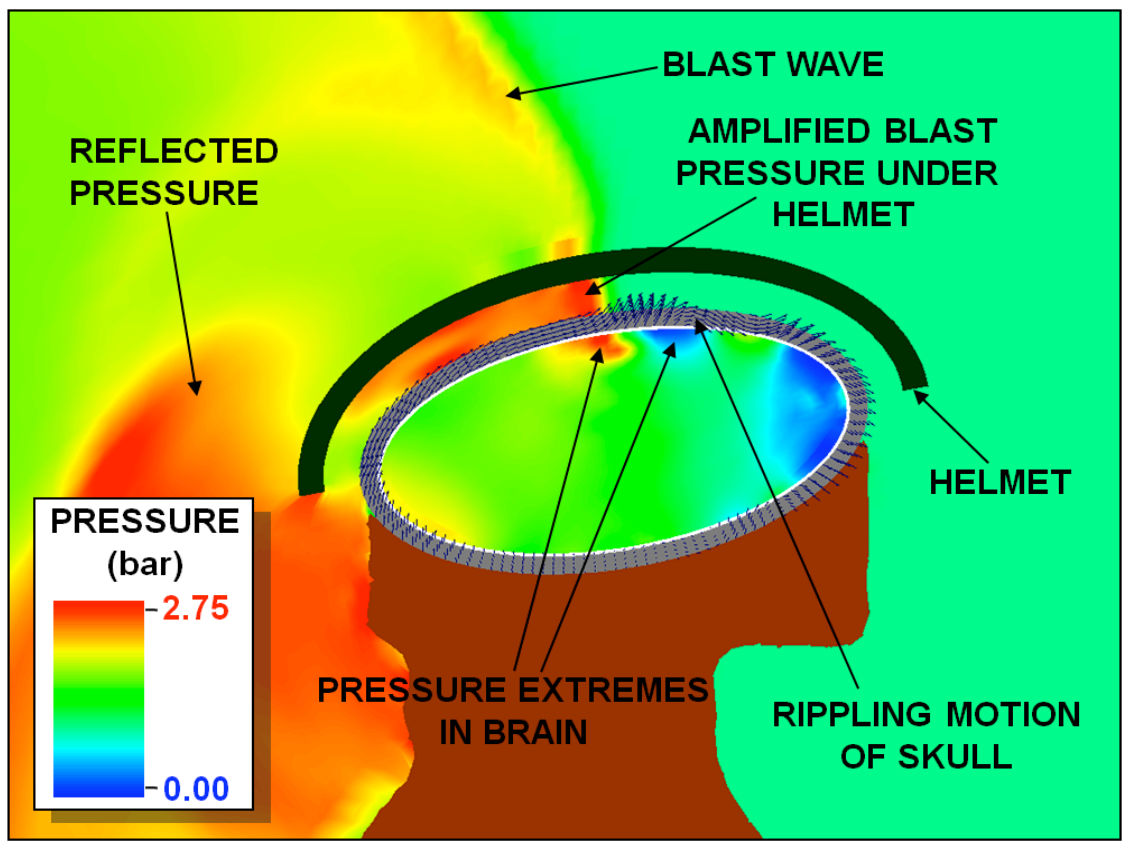

**Figure 4** –Amplification of the blast pressure and loads on the head due to "underwash" for a helmet without foam pads.

## Supplementary Methods

All simulations were conducted using ALE3D (*1*) in explicit dynamics mode, with most of the structural regions (the skull, CSF, brain, helmet shells, webbing, and impact foam) held Lagrangian and the other regions (air, detonation products, ACH foam, and body form) allowed to relax to prevent mesh entanglement. All elements except those used for the nylon webbing were 3D linear reduced integration elements; the webbing was represented with linear shell elements overlaid onto the advecting air mesh between the helmet and the head.

In the blast simulations, the ground was represented as a reflecting plane, as were all symmetry planes. Other boundaries in the blast simulations used an "outflow" boundary condition allowing material to pass out of the simulation space. All materials were initialized to zero initial velocity and 1 bar of ambient pressure. In the impact simulation, the head and surrounding "helmet" were initialized to a constant initial velocity. The object against which the head was impacted was a boundary configured to act as a frictionless rigid wall.

The explosive was detonated instantaneously at the start of the analysis and the products were described using the JWL equation of state (*2*). The air was described by a gamma-law gas equation of state ($\gamma$ = 1.4). Predicted blast pressure histories were validated against tabulated experimental blast data at distances ranging from 1.5 to 6.0 m.

The base case skull was modeled as an isotropic linear elastic hollow ellipsoid with semi-axes of 10, 8, and 5 cm and a constant 7 mm thickness, inclined at a 10° angle. The CSF was modeled as a water layer 1.3 mm thick. These dimensions are typical of an adult male. In all cases except case (*iii*), the CSF tensile strength is capped at one bar below atmospheric pressure; when this hydrostatic stress is reached, subsequent volumetric expansions cause no additional increase in hydrostatic stress.

Reported material properties for human cranial bone vary widely. The skull is frequently treated as a linear elastic structure, and some viscoelastic properties of the skull measured at lower rates (*3*) justify this approximation for our simulations, due to the short duration of the blast loading relative to the viscoelastic relaxation time. However, cranial bone properties measured at higher rates have indicated shorter viscoelastic relaxation times (*4*), so we conducted case (*vi*) using viscoelastic properties for the skull derived to fit the reported data, to ensure that our results were not sensitive to skull viscoelasticiy. Although an actual skull is a sandwich structure composed of stiff cortical bone on the faces and soft trabecular bone in the middle, we used a homogeneous skull for most of our simulations, with properties determined by a weighted average (according to the relative thicknesses) of the cortical and trabecular properties: density $\rho$ = 1.7 g/cm$^3$, Young's modulus $E$ = 9 GPa, Poisson's ratio $\nu$ = 0.229 for the elastic case, and instantaneous shear modulus $G_0$ = 4.4 GPa, quasi-static shear modulus $G_\infty$ = 2.3 GPa, bulk modulus $K$ = 10.5 GPa, decay factor $\beta$ = 2,237 s$^{-1}$ for case (*vi*). Higher fidelity simulations that explicitly modeled cortical and trabecular bone layers with the appropriate elastic properties ($E^{cortical}$ = 15 GPa, $\nu^{cortical}$ = 0.24, $\rho^{cortical}$ = 2.0 g/cm$^3$, $E^{trabecular}$ = 1 GPa, $\nu^{trabecular}$ = 0.22, $\rho^{trabecular}$ = 1.3 g/cm$^3$) produced no substantive difference in the results. Elastic properties for the skull were given by Horgan (*5*); viscoelastic properties were determined from the data given by Wood (*4*).

The brain tissue and face/body were modeled using a material with a linear equation of state and a viscoelastic strength law. For the brain (base case), $\rho$ = 1.04 g/cm$^3$, $G_\infty$ = 6.95 kPa, $G_0$ = 37.5 kPa, $K$ = 2.19 GPa, $\beta$ = 700 s$^{-1}$. These properties were averages of properties for grey and white matter given by Zhou *et. al* as reported by Horgan (*5*). For case (*iv*), we used $\rho$ = 1.04 g/cm$^3$, $G_\infty$ = 168 kPa, $G_0$ = 528 kPa, $K$ = 0.128 GPa, and $\beta$ = 35 s$^{-1}$ as given by Ruan *et. al* (*6*). For case (*v*), all properties were the same as case (*iv*) except that $K$ = 2.19 GPa. The face/body density was chosen to be representative of average body density, $\rho$ = 1.04 g/cm$^3$; the bulk modulus was that of water, and the shear response was arbitrarily chosen to be sufficiently stiffer than the brain so that it would keep its shape under blast loading.

The Kevlar helmet was a hollow hemiellipsoid with a constant thickness and offset from the skull as described by Reynosa (*7*), with transversely isotropic elastic material properties given by Aare and Keliven (*8*). The suspension systems are shown in Supplementary Figure 1. The ACH foam pad geometry was measured from foam pads removed from an ACH (*9*). The foam was modeled with a linear bulk response and a viscoelastic strength law. The properties were measured approximately at LLNL using low-rate compression tests and acoustic tests: $\rho = 0.136$ g/cm$^3$, $G_\infty = 20.1$ kPa, $G_0 = 2.0$ MPa, $K = 1.3$ MPa, $\beta = 100$ s$^{-1}$. However, because of the difficulty in measuring the properties of soft foam, these values are uncertain. Consequently, simulations were conducted over a range of foam stiffnesses, as described in the text. The nylon webbing geometry was measured from a PASGT helmet. Its stiffness was estimated from the modulus of nylon and the effective stiffness of a plain woven structure as described by King (*10*).

## Supplementary References

## Supplementary Figure

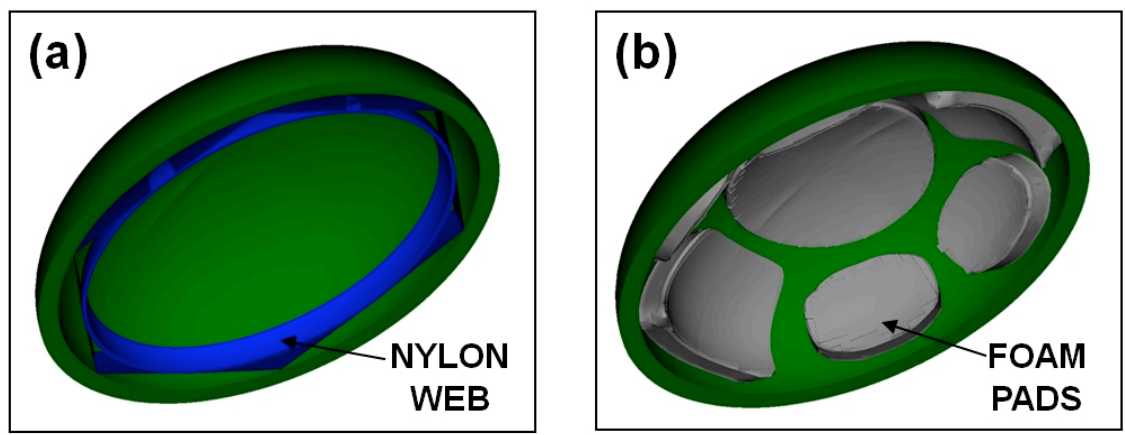

**Supplementary Figure 1** – Helmet suspensions modeled:

(a) PASGT type nylon web

(b) ACH type foam pads